\documentclass[secnumarabic,amssymb, nobibnotes, aps, prd, 11pt, abstract]{revtex4-2}
\usepackage{tocloft}
\usepackage[dvips]{graphicx}
\usepackage{amsmath}
\usepackage{subcaption}
\usepackage{hyperref}
\usepackage{filecontents}
\hypersetup{colorlinks,linkcolor={red},citecolor={blue},urlcolor={red}}
\usepackage{setspace}
\usepackage{xcolor}
\pagecolor{white}
\usepackage{float}

\usepackage{multirow}
\color{black}

\begin{document}

\singlespacing

	\title{Quasi-Periodic Oscillations and Parameter Constraints in ModMax Black Holes}%

	\author{Mozib Bin Awal$^1$}
	
	\email{$mozibawal@gmail.com$}
	
	\author{Bidyut Hazarika$^1$}
	
	\email{$rs_bidyuthazarika@dibru.ac.in$}

	\author{Prabwal Phukon$^{1,2}$}
	\email{$prabwal@dibru.ac.in$}
	
	\affiliation{$^1$Department of Physics, Dibrugarh University, Dibrugarh, Assam,786004.\\$^2$Theoretical Physics Division, Centre for Atmospheric Studies, Dibrugarh University, Dibrugarh,Assam,786004.\\}

	\begin{abstract}
	
We analyze the impact of ModMax parameter on the dynamics of test particles around black holes and its effect on the characteristics of Quasi-Periodic Oscillations (QPOs). The effect of the ModMax parameter $\eta$ is studied using the effective potential, angular momentum and the energy of the circular orbits of the test particles. Our analysis shows that increasing $\eta$ brings about a continuous transition from the Reissner–Nordström (RN) regime toward the Schwarzschild limit, accompanied by noticeable modifications in the Innermost Stable Circular Orbit (ISCO) and the corresponding Keplerian frequencies. We also explore the dependence of QPO radii on the ModMax parameter $\eta$ within the framework of the PR, RP, WD, and ER models. Finally, to place observational constraints, we perform a Markov Chain Monte Carlo (MCMC) analysis using QPO data from a range of black hole sources spanning stellar-mass, intermediate-mass, and supermassive scales.
 
\end{abstract}
	
	\maketitle
	
\section{Introduction}\label{sec1}
Black holes are among the most mysterious and fascinating objects in this entire universe. It is a profound outcome of Einstein's General Theory of Relativity which describes gravity as curvature of spacetime. Over the decades, General Relativity (GR) has passed several experimental verifications. However, the most compelling verification of GR came from LIGO collaboration which detected the ripples in spacetime caused due to the merger of black holes in the form of gravitational waves \cite{ligo}. Following this, another major breakthrough was the imaging of the supermassive black hole in the galaxy M87 and then of the black hole Sagittarious A* situated at our own galaxy Milkyway by the Event Horizon Telescope \cite{m87a, m87b, m87c, m87d, m87e, m87f}. These images reveal a structure having a central shadow surrounded by a luminous photon ring,the study of which provides valuable insight into the underlying gravitational framework and the fundamental nature of black holes \cite{Shadow1,Shadow2,Shadow3,Shadow4}. More studies suggested that the structure of black hole shadows and the properties of photon spheres provide powerful means to probe and constrain deviations from general relativity, thereby serving as essential tools for testing alternative theories of gravity \cite{Shadow5,Shadow6,Shadow7,Shadow8,Shadow9,Shadow10,Shadow11,Shadow12}.

Studying particle dynamics in the vicinity of black holes have proved to be very essential for probing their geometric as well as physical features. Over the years, numerous studies have explored the trajectories of both massive and massless particles across different parameterized black hole geometries \cite{2,3,4,5,6,7,8,9,10,11,12}. Many studies have focused on the orbital and epicyclic frequencies in axially symmetric and stationary spacetimes, highlighting their importance in exploring particle dynamics around black holes \cite{11}. Early studies provided exact analytical solutions for geodesic motion, forming the foundation for subsequent, more sophisticated analyses \cite{13}. Later works expanded these results to encompass the motion of charged test particles in spacetimes influenced by both electric and magnetic fields \cite{14,15}. Recent findings indicate that the combined influence of electric charge and external magnetic fields in Reissner–Nordström spacetime can reproduce the characteristics of magnetically charged black holes \cite{16}, adding further intricacy to particle motion in such configurations.

The observation of Quasi-Periodic Oscillations (QPOs) in the X-ray emissions of black holes and neutron stars has emerged as powerful tools for exploring the physics of strong gravitational fields. These oscillations are characterised by their nearly periodic variations in intensity and are believed to occur form fundamental mechanisms such as accretion disk dynamics and relativistic gravitational effects. Particularly, the detection of the twin-peak QPOs in certain systems has motivated extensive studies aimed at identifying their physical origin, often associated with resonant or oscillatory modes within the accretion disk. Lately, the growing demand for more sophisticated theoretical frameworks supported by high precision observational data has become more apparent. And with the very first detection of QPOs via spectral and timing  analyses in X-ray binaries \cite{17}, the phenomenon has been extensively studied from both theoretical and observational perspectives. Notably, the models describing particle motion in curved spacetime has proven to be particularly promising where the observed oscillatory behaviour is attributed to charged test particles, shaping the structure and evolution of the accretion flow \cite{18,19,20,21,22,23,24,25,26,27,28,29,30,31,32}. Recent works have numerically investigated the physical mechanisms underlying QPO generation in black hole systems by solving general relativistic hydrodynamic equations \cite{33} for spacetimes including Kerr and hairy black holes. These simulations indicate that the plasma perturbations during accretion can trigger spiral shock waves which could be associated with QPO activity \cite{34,35,36}. Similar studies based on Bondi–Hoyle–Lyttleton accretion suggest that shock cones formed in intense gravitational fields can yield characteristic QPO frequencies \cite{37,38,39,40,41}. Such models have successfully accounted for observed QPOs in sources like GRS 1915+105 \cite{42} and have also provided predictions for QPO features in the vicinity of supermassive black holes such as M87* \cite{43}. Extensive research has focused on the motion of test particles and their corresponding QPOs around black holes, for example one may visit Refs. \cite{c1,c2,c3,c4,c5,c6,c7} for representative studies. In particular, Ref. \cite{c2} analyzed black holes derived from nonlinear electrodynamics in relation to observed QPOs, emphasizing regular rotating configurations.

Nonlinear electrodynamics (NED) has emerged as an effective framework for generating regular and physically consistent field configurations in curved spacetime \cite{44}. Unlike linear Maxwell theory, nonlinear electrodynamics characterised by gauge invariant Lagrangians dependent on the electromagnetic invariant \( F = F_{\mu\nu}F^{\mu\nu} \) can yield stress-energy tensors exhibiting symmetries that imitate vacuum behaviour under radial boosts. Notably, among these theories, Born-Infeld (BI) electrodynamics has attracted considerable attention for its appearance in the low-energy regime of string theory. Furthermore, several other NED formulations \cite{45,46,47,48,49} share desirable features with the Born-Infeld theory, such as finite electric field strength at the origin and finite total electrostatic energy. Another nonlinear theory worth mentioning in this regard is the Euler Heisenberg Nonlinear Electrodynamics (EH NLED) theory \cite{Salazar:1987ap}. Both Euler Heisenberg and Born Infeld NLED preserve the electric-magnetic duality invariance. More recently a different non linear electrodynamics theory endowed with two symmetries was proposed \cite{Bandos:2020jsw}. This particular framework was characterised by the dimensionless parameter $\eta$ and for $\eta=0$ it reduces to the standard Maxwell theory. This modified version of electrodynamics, commonly referred to as the ModMax NLED theory, has stimulated extensive research in several directions, including the study of classical solutions \cite{Banerjee:2022sza} as well as supersymmetric extensions \cite{Escobar:2022lpu}. In what follows, we will review the features of the ModMax black hole adopted in this study. It is important to emphasize that NED effects, including those described by ModMax theory, are far too weak to be experienced in everyday or laboratory scale electromagnetic phenomena. Such nonlinear corrections become significant only in the presence of extremely strong electromagnetic and/or gravitational fields, as encountered in the vicinity of black holes. In this sense, black holes serve as natural laboratories for probing and testing nonlinear electrodynamic theories.

We start from the action of the ModMax black hole \cite{Sekhmani} \begin{equation}\label{eq1}
\mathcal{I}=\frac{1}{16\pi}\int d^4x\sqrt{-g}\left(\mathcal{R}-4\mathcal{L}'\right)
\end{equation}here $\mathcal{R}$ is the Ricci scalar, $g$ denotes the determinant of the metric tensor $g_{\mu\nu}$ and $\mathcal{L}'$ denotes the ModMax Lagrangian. This Lagrangian can also be written as \cite{Kosyakov} \begin{equation}\label{eq2}
\mathcal{L}'=\frac{1}{2}\left(\mathcal{S}\cosh{\eta}-\sqrt{\mathcal{S}^2+\mathcal{P}^2}\sinh{\eta}\right)
\end{equation} where $\eta$ is a dimensionless quantity intrinsic to the ModMax theory.  $\eta$ can be interpreted as an effective coupling parameter that  characterizes the strength of nonlinear deviations from standard Maxwell electrodynamics. $\mathcal{S}$ and $\mathcal{P}$ are two real and pseudo-scalars respectively. The metric of a spherically symmetric charged ModMax black hole can be written as  \begin{equation}\label{eq3}
ds^2=-f(r)^2dt^2+\frac{1}{f(r)^2}dr^2+r^2\left(d\theta^2+r^2\sin^2\theta d\phi^2\right)
\end{equation} where the metric function is given by \begin{equation}\label{eq4}
f(r)=1-\frac{2 M}{r}+\frac{e^{-\eta } Q^2}{r^2}
\end{equation}
There are a few noteworthy features that are arise in the extremal limits of the parameter $\eta$ which hold particular significance in the context of the present study. In the limit $\eta \to 0$, the metric function of ModMax black hole reduces to that of a Reissner–Nordström (RN) solution indicating that for sufficiently smaller values of $\eta$, the ModMax geometry should closely resemble to that of the RN black hole. In the other extremal limit, when $\eta \to \infty$, the metric simplifies to the usual Schwarzschild solution. So, in the large $\eta$ limits, the ModMax spacetime should mimic the spacetime of a Schwarzschild black hole. 

The primary motivation of our work is to examine whether the characteristic behaviour of the spacetime metric observed in the extremal limits of the ModMax parameter $\eta$ is manifested in the properties of quasi-periodic oscillations. In particular, we aim to explore whether similar trends appear in the QPO frequencies as $\eta\to 0$ and $\eta\to \infty$, and to assess how the ModMax parameter influences these features. Ultimately, our goal is to identify possible signatures of nonlinear electrodynamics imprinted in the QPO profiles by assessing these characteristics.

In this work, we investigate the motion of neutral test particles in the vicinity of a static, spherically symmetric,  charged black hole arising from nonlinear electrodynamics coupled to general relativity. Our main objective is to examine how the ModMax parameter $\eta$ influences the quasi-periodic oscillations associated with such black holes. Furthermore, to investigate the influence of the ModMax parameter $\eta$ analyze the variation of QPO radii across different phenomenological models. These include the Relativistic Precession (RP), Warped Disk (WD), and Epicyclic Resonance (ER) models. Moreover, we perform a Markov Chain Monte Carlo (MCMC) analysis using observational QPO data from a range of black hole sources spanning stellar-mass, intermediate-mass, and supermassive regimes. The resulting posterior distributions provide consistent constraints on the parameter $\eta$ indicating that the effects of nonlinear electrodynamics imprint discernible signatures on the QPO spectra of charged black holes across all mass scales.

\section{Particle dynamics around modmax black holes}
\subsubsection{Equations of Motion}
We now investigate the motion of electrically neutral test particles in the spacetime of a charged black hole governed by nonlinear electrodynamics. The dynamics of such particles are described by the Lagrangian, \begin{equation}\label{5}
 L_p = \frac{1}{2} m g_{\mu\nu} \dot{x}^\mu \dot{x}^\nu
\end{equation}
where $m$ is the mass of the particle. The dot denotes differentiation with respect to the proper time $\tau$. It is to be noted that \( x^\mu(\tau) \) represents the worldline of the particle, parametrized by the proper time $\tau$ while the corresponding four-velocity is defined as \( u^\mu = \frac{dx^\mu}{d\tau} \). In a spherically symmetric spacetime, two Killing vectors arise from the invariance of the metric under time translations and spatial rotations, expressed as  \( \xi^\mu = (1, 0, 0, 0) \) and \( \eta^\mu = (0, 0, 0, 1) \), respectively. So, the constant of motion associated with the total energy anf angular momentum can be written as, \begin{align}
\mathcal{E}&=-g_{tt}  ~\Dot{t}, \nonumber\\
\mathcal{L}&=g_{\phi \phi}~ \Dot{\phi}.
\label{energy}
\end{align} In the above equation, $\mathcal{E}$ and $\mathcal{L}$ are the energy and angular momentum per unit mass respectively. We can then formulate the equation of motion for the test particle using the following normalisation condition \begin{align}\label{normal}
    g_{\mu\nu} u^\mu u^\nu = \delta,
\end{align}
where $\delta = 0$ corresponds to the geodesic motion associated to massless particles and $\delta=\pm1$ corresponds to that of massive particles. In particular, $\delta=+1$ represents spacelike geodesics, while $\delta=-1$ corresponds to timelike geodesics. For massive particles, the motion is thus governed by timelike geodesics of the spacetime, and the corresponding equations can be derived using Eq.(\ref{normal}). By combining Eqs.~(\ref{energy}) and (\ref{normal}), the equation of motion on a constant plane can be written in the form\begin{align}
    \dot{r}^2=\mathcal{E}^2+g_{tt}\Big(1+\frac{{\mathcal{L}^2}}{r^2 }\Big)
\end{align}
For a static, spherically symmetric spacetime if we consider the particle to begin its motion in the equatorial plane, its motion will continue in that plane throughout its trajectory. Once we restrict the motion in the equatorial plane by considering  $\theta = \frac{\pi}{2}$ and $\dot{\theta} = 0$, the radial equation of motion can be expressed as \begin{align}
    \dot{r}^2 = \mathcal{E}^2 - V_\text{eff},
\end{align}
For circular orbits, we have $\dot{r}=0$ and $\ddot r=0$ and we have, \begin{align}
 V_\text{eff}=\mathcal{E}^2
     \label{cosnt}
\end{align}
Here $V_\text{eff}$ is the effective potential that governs the radial motion in the equatorial plane. It is given by \begin{align}
    V_\text{eff} = f(r) \left(1 + \frac{\mathcal{L}^2}{r^2}\right).
\end{align}
\begin{figure}[h!]
\centering
\includegraphics[width=0.5\linewidth]{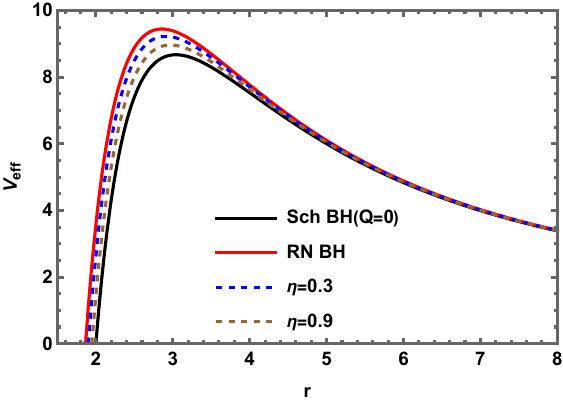}
\caption{Dependence of the effective potential $V_\text{eff}$ on the radial coordinate $r$ for different values of ModMax parameter $\eta$. Here $Q=0.5$ } 
\label{fig1}
\end{figure}
Figure \ref{fig1} shows the dependence of the effective potential for a neutral particle with respect to the radial coordinate $r$ for different values of the ModMax parameter $\eta$. Here we also show the comparison of the results with those for the Reissner-Nordström and the Schwarzschild black hole ($Q=0$). We observe that for neutral test particles, when the charge of the black hole $Q$ is kept fixed, increasing the value of $\eta$ results in the decrease of the maximum of the effective potential. If we keep on increasing $\eta$, the profile of the effective potential approaches that of the Schwarzschild black hole depicted by the black curve in Figure \ref{fig1}. Stable circular orbits arise at the minima of the effective potential, where a particle can maintain equilibrium without significant radial displacement. Conversely, the maxima correspond to unstable circular orbits, in which even small perturbations can drive the particle away from its trajectory. The location and depth of these extrema depend on the black hole parameters $Q$ and the ModMax parameter $\eta$. These two together shape the spacetime geometry and influence the resulting particle motion.

Now using $\dot{r}=0$ and equation \ref{cosnt}, we can write down the expression for specific energy and angular momentum for circular orbits 
\begin{equation}
\mathcal{L}^2=\frac{\sqrt{M r-e^{-\eta } Q^2}}{\sqrt{\frac{-3 M r+2 e^{-\eta } Q^2+r^2}{r^2}}}
\end{equation}
\begin{equation}
\mathcal{E}^2=\frac{e^{-\eta } \left(e^{\eta } r (r-2 M)+Q^2\right)^2}{r^2 \left(e^{\eta } r (r-3 M)+2 Q^2\right)}
\end{equation}

\begin{figure}[h!]
\centering
\includegraphics[height=4cm,width=0.32\linewidth]{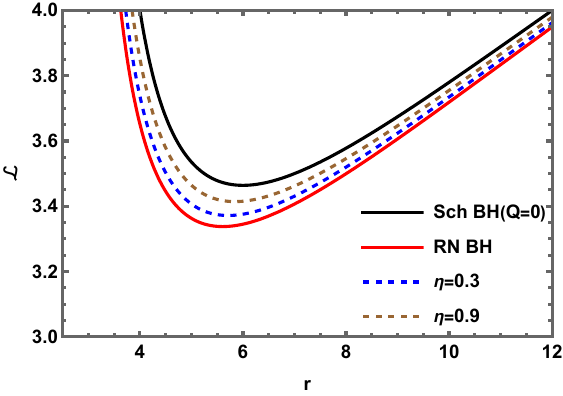}  % Adjust width to fit three figures
\includegraphics[height=3.9cm,width=0.32\linewidth]{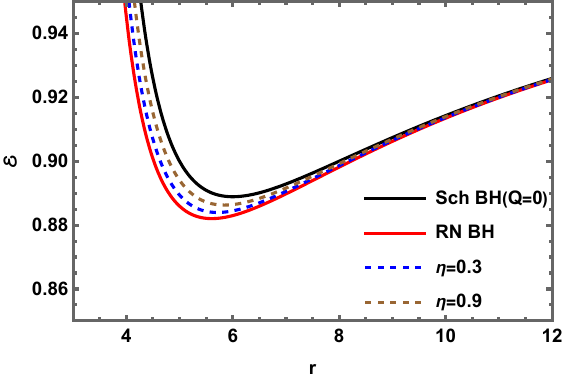}
\includegraphics[height=4cm,width=0.32\linewidth]{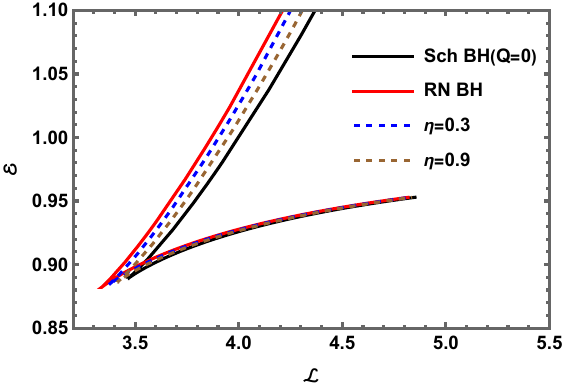}
\vspace{-0.3cm}
\caption{The radial dependence of specific angular momentum and energy  for circular orbits for different values of NED parameter b.  Here, we have considered $Q = 0.5$.}
\label{fig2}
\end{figure}

In Figure \ref{fig2}, we show the variation of specific energy and angular momentum for circular orbits with respect to the radial coordinate $r$ highlighting the influence of the ModMax parameter $\eta$. In the left panel the radial dependence of the specific angular momentum is shown. Here also, the solid black line corresponds to Schwarzschild black hole, while the solid red curve corresponds to Reissner-Nordström black hole. The blue and red dashed curves illustrate the modifications introduced by the ModMax parameter. We observe that for large $\eta$ the behaviour of $\mathcal{L}$ converges towards Schwarzschild black hole and for smaller $\eta$ it resembles that of the Reissner-Nordström case. In the middle panel we show the variation of the specific energy with respect to the radial coordinate using the same colour coding as before. 

The energy profile displays a distinct minimum, corresponding to the most tightly bound circular orbit. Relative to the Schwarzschild and Reissner-Nordström black holes, the presence of the ModMax parameter alters the energy associated with stable orbits, modifying both the depth and position of this minimum in a similar manner to that of the specific angular momentum. The right panel illustrates the correlation between $\mathcal{L}$ and $\mathcal{E}$ offering insights into orbital stability and their dependence on the ModMax parameter. It is observed that for larger values of $\eta$ both the specific angular momentum $\mathcal{L}$ and energy $\mathcal{E}$ of circular orbits decrease, implying that the orbits become more tightly bound. From a stability perspective, lower $\mathcal{L}$ and  $\mathcal{E}$ values indicate that a particle requires less energy to remain in orbit, causing the stable circular orbits to move outward. Nonetheless, if this reduction becomes pronounced, the stable region may contract, increasing the likelihood of orbital instabilities. Clear distinctions emerge among the Schwarzschild, Reissner-Nordström, and ModMax configurations, with the latter showing noticeable deviations driven by the parameter $\eta$.

\subsection{Innermost Stable Circular Orbits (ISCO)}
By solving $V_{\text{eff}} = 0$ for $r$, we get the location where the effective potential shows extremal behaviour. In the case of the innermost stable circular orbit (ISCO), an additional condition is required to guarantee stability against small perturbations. The effective potential $V_{\text{eff}}$ which governs the particle’s radial motion, must satisfy that both its first and second derivatives with respect to $r$ vanish, i.e., $V'_{\text{eff}} = 0$ and $V''_{\text{eff}} = 0$. These criteria together define the exact position of the ISCO, ensuring that the orbit remains circular and is marginally stable against radial disturbances.\begin{figure}[!h]
\centering
\includegraphics[width=0.5\linewidth]{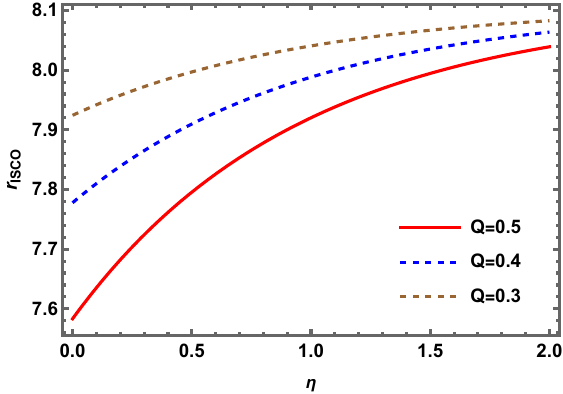}
\vspace{-0.3cm}
\caption{Dependence of the ISCO radius on the ModMax parameter $\eta$ } 
\label{fig3}
\end{figure}
In Figure \ref{fig3} we present the variation of the ISCO radius with respect to the ModMax parameter for different values of charge parameter $Q$. The ISCO radius is observed to be increasing with increasing $\eta$ for all values of $Q$ implying that the presence of NLED forces the ISCO to shift outwards from the black hole. Moreover, for a given value of $\eta$, the ISCO radius shows an approximately linear increase with the charge parameter $Q$. This behaviour is consistent with expectations since, in the limit $Q \to 0$ the normalized ISCO radius approaches a constant value of $\frac{r_{\text{ISCO}}}{M} = 6$ corresponding to the Schwarzschild black hole limit.

\section{Fundamental frequencies}
In this section, we calculate the fundamental frequencies that describe the motion of a particle near a ModMax black hole. Specifically, we analyze the Keplerian frequency, as well as the radial and vertical epicyclic frequencies corresponding to small perturbations around circular orbits.

\subsection{Keplerian Frequencies}
\begin{figure}[!h]
\centering
\includegraphics[width=0.5\linewidth]{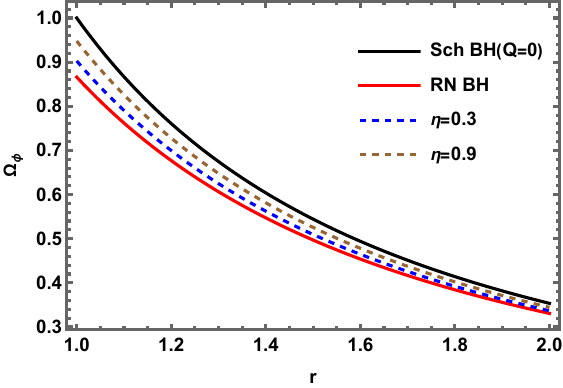}
\vspace{-0.3cm}
\caption{Variation of angular frequency $\Omega_\phi$ with respect to $r$}
\label{fig4}
\end{figure}

The angular velocity of a test particle moving around a black hole, as observed from infinity, is referred to as the orbital or Keplerian frequency, represented by $\Omega_\phi$. It is defined through the relation $\Omega_\phi = \frac{d\phi}{dt}$. Based on this definition, the general form of the orbital frequency in a static and spherically symmetric spacetime can be expressed as \cite{50} \begin{align}
    \Omega_\phi = \sqrt{\frac{-\partial_r g_{tt}}{\partial_r g_{\phi\phi}}} = \sqrt{\frac{f'(r)}{2r}}.
\end{align}
For the ModMax black hole considered in our study, this expression comes out to be \begin{align}
\Omega_\phi = \sqrt{\frac{M r-e^{-\eta } Q^2}{r^4}}
\end{align}
Substituting $Q=0$ or $\eta\to \infty$, we find the same angular velocity as that of the pure Schwarzschild case \cite{c7}, i.e.,\begin{align}
\Omega_\phi = \sqrt{\frac{M}{r^3}}.
\end{align}
Again in the limit $\eta\to 0$, we arrive at the angular velocity of the RN case, which is \begin{align}
\Omega_\phi = \sqrt{\frac{M}{r^3}-\frac{Q^2}{r^4}}.
\end{align}
To express the angular frequency in physical units of Hertz (Hz), the following conversion relation is used, 
\begin{align}
    \nu_\phi = \frac{c^3}{2\pi G M}\Omega_\phi
\end{align}
We present the variation of the Keplerian frequency $\Omega_\phi$ with respect to the radial coordinate $r$ in Figure \ref{fig4}. We show the angular frequency dependence for different values of the ModMax parameter $\eta$ for fixed value of $Q$ and also give the comparison with the Schwarzschild and RN case. It is observed that the orbital frequency $\Omega_\phi$ decreases with increasing radial distance $r$. 
The presence of charge and NLED effects further suppresses $\Omega_\phi$ relative to the Schwarzschild case. Increasing the ModMax parameter $\eta$ causing an additional decrease in the orbital frequency, emphases its impact on particle dynamics. In agreement with earlier results, as $\eta$ becomes larger, the frequency profile approaches that of the Schwarzschild black hole, whereas smaller values of $\eta$ make it resemble the Reissner–Nordström (RN) case.

\subsection{Harmonic Oscillations}
We now investigate the fundamental frequencies corresponding to the oscillatory motion of test particles orbiting a ModMax black hole. The associated  characteristic frequencies, namely, the radial and vertical (or latitudinal) components are derived by introducing small perturbations around the equilibrium circular as $r \rightarrow r_0 + \delta r$ and $\theta \rightarrow \theta_0 + \delta \theta$. We then Taylor expand the effective potential around the circular orbit $(r_0, \theta_0)$ which gives, \begin{align}
V_{\text{eff}}(r, \theta) &= V_{\text{eff}}(r_0, \theta_0) + \delta r \left. \frac{\partial V_{\text{eff}}}{\partial r} \right|_{r_0, \theta_0} + \delta \theta \left. \frac{\partial V_{\text{eff}}}{\partial \theta} \right|_{r_0, \theta_0} \nonumber + \frac{1}{2} \delta r^2 \left. \frac{\partial^2 V_{\text{eff}}}{\partial r^2} \right|_{r_0, \theta_0} + \frac{1}{2} \delta \theta^2 \left. \frac{\partial^2 V_{\text{eff}}}{\partial \theta^2} \right|_{r_0, \theta_0} \nonumber \\
&+ \delta r \delta \theta \left. \frac{\partial^2 V_{\text{eff}}}{\partial r \partial \theta} \right|_{r_0, \theta_0} + \mathcal{O}(\delta r^3, \delta \theta^3).
\label{exp}
\end{align}
Enforcing the conditions for circular motion and orbital stability, one finds that only the second-order derivatives of the effective potential remain relevant, resulting in harmonic oscillator-type equations in the equatorial plane for both radial and vertical perturbations. These oscillations, as perceived by a distant observer, are expressed as \cite{Bambi2017book} \begin{align}
    \frac{d^2\delta r}{dt^2}+\Omega^2_r\delta r=0,\  \frac{d^2\delta\theta }{dt^2}+\Omega^2_\theta \delta\theta=0,
\end{align}
where
\begin{align}
 \Omega_r^2=-\frac{1}{2g_{rr}\Dot{t}^2}\partial_r^2 V_\text{eff}(r,\theta)\Big\arrowvert_{\theta=\pi/2},
\end{align}
\begin{align}
        \Omega_\theta^2=-\frac{1}{2g_{\theta\theta}\Dot{t}^2}\partial_\theta^2 V_\text{eff}(r,\theta)\Big\arrowvert_{\theta=\pi/2},
\end{align}
represent the frequencies corresponding to the radial and vertical oscillations, respectively. The expressions for the frequencies of radial and vertical oscillations for ModMax black holes are 
\begin{align}
 \Omega_r=\sqrt{\frac{e^{-2 \eta } \left(e^{\eta } r (r-3 M)+2 Q^2\right)^2 \left(Q^2 r \left(e^{\eta } (8 M-3 r)+M+3 r\right)-e^{\eta } M r^2 (6 M-r)-4 Q^4\right)}{r^6 \left(r (r-3 M)+2 Q^2\right) \left(e^{\eta } r (r-2 M)+Q^2\right)}}
\end{align}
\begin{align}
        \Omega_\theta= \Omega_\phi=\sqrt{\frac{M r-e^{-\eta } Q^2}{r^4}}
\end{align}
Again, to express these frequencies in physical units (Hertz), we use the following conversion relation \begin{align}
    \nu_i = \frac{c^3}{2\pi G M} \cdot \Omega_i
\end{align}

\section{quasiperiodic oscillation model and orbits}
\subsection{QPO Models}
In this section, we analyze the characteristics of twin-peak quasi-periodic oscillations (QPOs) around black holes governed by nonlinear electrodynamics and compare them with the corresponding behaviours in Schwarzschild and Reissner-Nordström geometries. The upper ($\nu_U$) and lower ($\nu_L$) QPO frequencies are formulated as functions of the radial coordinate and the black hole parameters, following several well-known theoretical frameworks for QPO modeling \cite{51}. The study incorporates multiple QPO models to capture these dynamics comprehensively.
\begin{itemize}
 \item \textbf{Parametric resonance (PR) models:}The characteristic $3{:}2$ ratio frequently observed in twin-peak high-frequency quasi-periodic oscillations (HFQPOs) from black hole and neutron star systems has inspired the interpretation that these phenomena originate from resonant interactions between distinct oscillatory modes within the accretion disk~\cite{pr1,pr2,pr3}. In this picture, small perturbations in the radial and vertical directions around stable circular orbits are treated as two decoupled harmonic oscillations, governed by the respective epicyclic frequencies $\nu_r$ and $\nu_\theta$.  

According to the parametric resonance model, relatively stronger radial perturbations can drive vertical oscillations through nonlinear coupling, since in thin accretion disks the radial displacement ($\delta r$) typically exceeds the vertical one ($\delta \theta$). A resonance occurs when the ratio of the radial to vertical epicyclic frequencies satisfies $\nu_r / \nu_\theta = 2/n$, where $n$ is a positive integer. For rotating black holes, where generally $\nu_\theta > \nu_r$, the most prominent resonance appears for $n = 3$, giving rise to the characteristic $3{:}2$ frequency ratio. Within this framework, the observable QPO frequencies are identified as  
\begin{equation}
\nu_U = \nu_\theta, \quad \nu_L = \nu_r .
\end{equation}

\end{itemize}
\begin{figure*}[t!]
\centering
\includegraphics[width=0.32\linewidth]{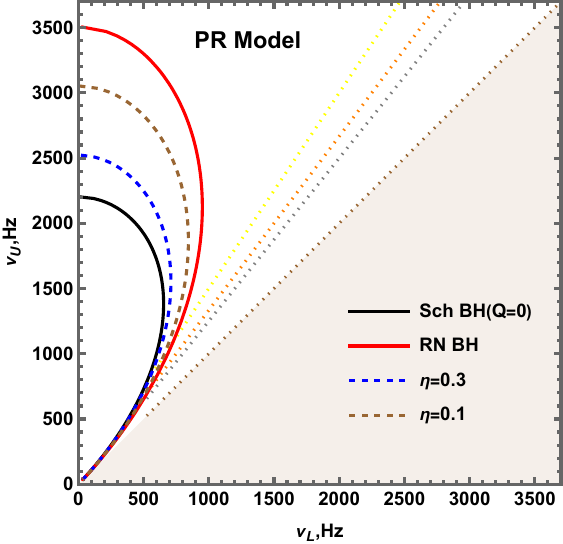}
\includegraphics[width=0.32\linewidth]{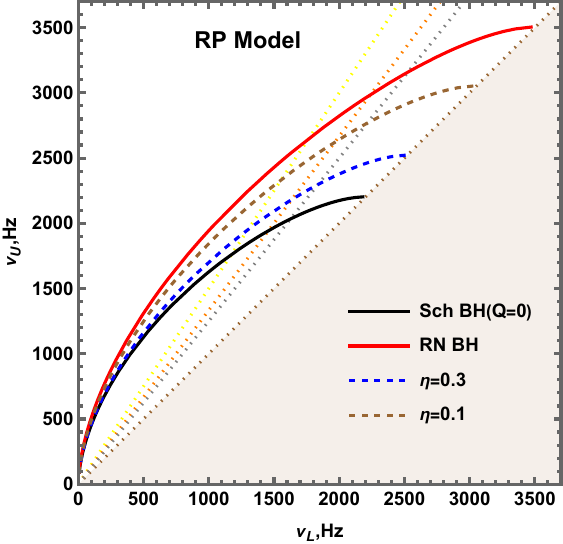}
\includegraphics[width=0.33\linewidth]{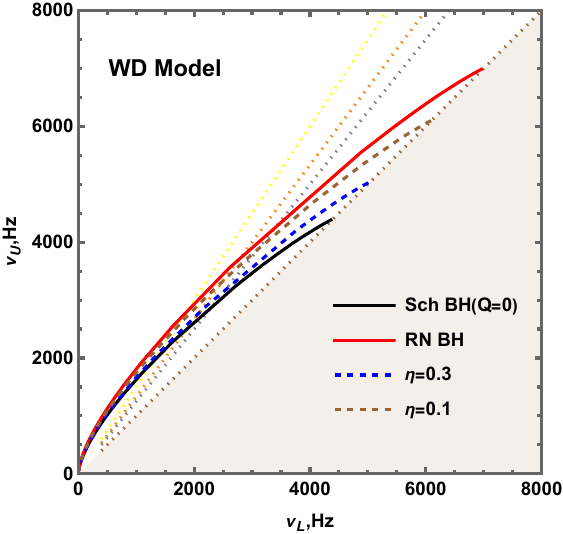}
\includegraphics[width=0.32\linewidth]{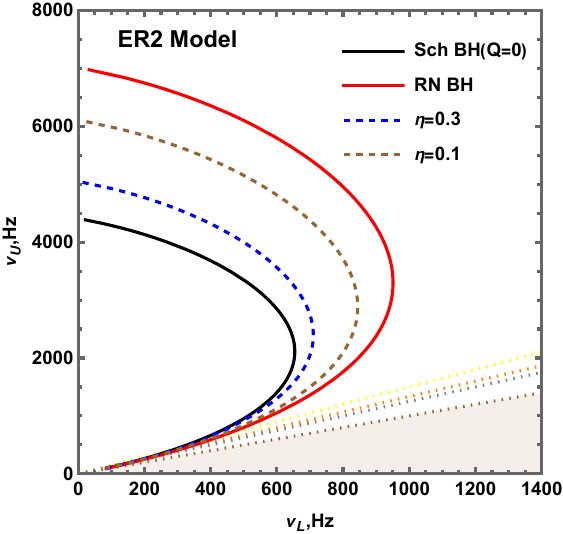}
\includegraphics[width=0.32\linewidth]{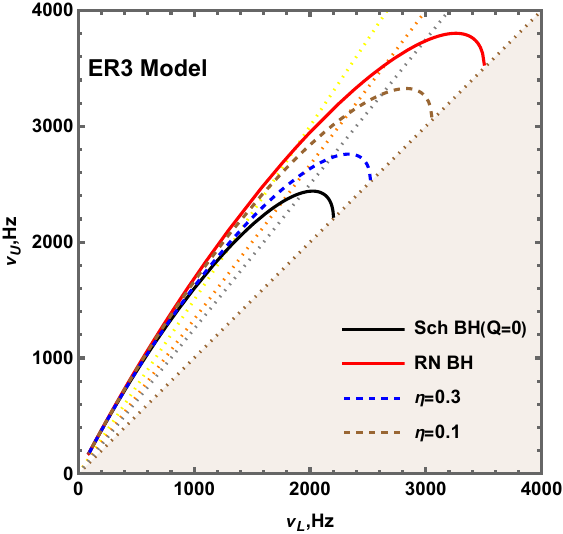}
\includegraphics[width=0.32\linewidth]{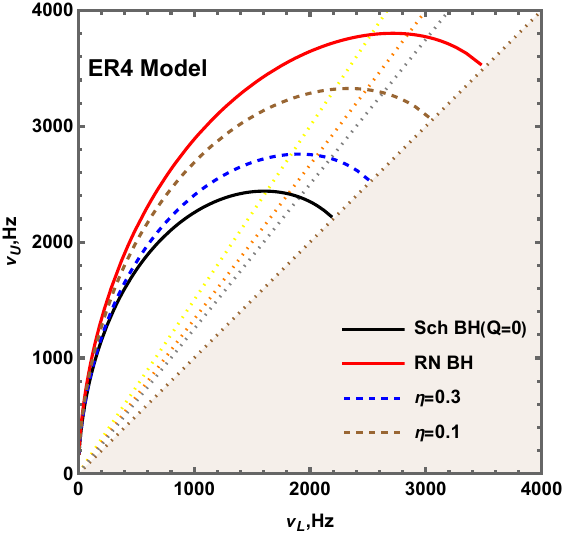}
\caption{The correlations between the upper ($\nu_U$) and lower ($\nu_L$) twin-peak QPO frequencies as analyzed under the PR, RP, WD, and ER2–ER4 models for Schwarzschild, RN, and ModMax black hole backgrounds.} 
\label{fig5}
\end{figure*}
\begin{itemize}
  \item \textbf{Relativistic Precession (RP) model:}In the framework of the relativistic precession (RP) model, the quasi-periodic oscillations (QPOs) detected in X-ray binaries are interpreted as a consequence of relativistic motion of accreting matter in the curved spacetime surrounding compact objects. Within this model, small clumps of plasma in the accretion disk are assumed to move along slightly eccentric and inclined geodesics close to the black hole. These mild deviations from perfect circular motion give rise to distinct oscillation frequencies that characterize the orbital dynamics.

The RP model connects the observed high-frequency QPOs with the fundamental coordinate frequencies of geodesic motion. Specifically, the upper frequency is associated with the orbital or Keplerian motion, $\nu_{U} = \nu_{\phi}$, whereas the lower frequency corresponds to the periastron precession frequency, $\nu_{L} = \nu_{\phi} - \nu_{r}$, where $\nu_{r}$ denotes the radial epicyclic frequency. Furthermore, the low-frequency QPOs observed in black hole systems are attributed to the nodal precession, characterized by $\nu_{\phi} - \nu_{\theta}$, which arises from vertical oscillations due to spacetime frame-dragging. In the Schwarzschild limit, this nodal precession frequency vanishes since $\nu_{\theta} = \nu_{\phi}$.

These frequency identifications naturally emerge from general relativistic corrections to orbital motion, including the effects of spacetime curvature and frame dragging. The RP model thus provides a geometrically grounded explanation for QPOs without invoking resonance phenomena or strong magnetic fields. Observational evidence from both neutron star and black hole systems, particularly the detection of harmonic structures and dominant even harmonics in power spectra, lends further support to this interpretation~\cite{StellaVietri1998, StellaVietri1999, MorsinkStella1999}.

\item\textbf{Epicyclic Resonance (ER) model:}The Epicyclic Resonance (ER) model provides a relativistic explanation for high-frequency quasi-periodic oscillations, interpreting them as manifestations of nonlinear couplings between the fundamental oscillation modes of matter in the accretion flow around compact objects. Within this framework, QPOs emerge from resonant interactions between the radial and vertical epicyclic motions of test particles following slightly perturbed geodesics in the accretion disk. In the present analysis, we consider three representative variants of the ER model, namely, the ER2, ER3, and ER4 model, each defined by specific resonant relations involving the orbital and epicyclic frequencies. The corresponding QPO frequency relations are expressed as follows,
 \begin{itemize}
        \item ER2: $\nu_U = 2\nu_\theta - \nu_r$, $\nu_L = \nu_r$,
        \item ER3: $\nu_U = \nu_\theta + \nu_r$, $\nu_L = \nu_\theta$,
        \item ER4: $\nu_U = \nu_\theta + \nu_r$, $\nu_L = \nu_\theta - \nu_r$.
    \end{itemize}
\end{itemize}

\begin{itemize}
\item \textbf{Warped Disk (WD) model:}
This model introduces a non-axisymmetric geometry for the accretion disk~\cite{wd1,wd2}, attributing high-frequency QPOs to nonlinear resonant interactions between the relativistically distorted, warped disk and its intrinsic oscillation modes. 
In this scenario, resonances arise through both horizontal and vertical couplings: the horizontal modes can excite both g-mode and p-mode oscillations, whereas the vertical interactions predominantly stimulate g-modes~\cite{wd2}. 
The emergence of such resonances is fundamentally connected to the non-monotonic radial dependence of the epicyclic frequency, which allows the disk to support multiple oscillation modes at specific radii. 
Within this framework, the characteristic QPO frequencies are given by~\cite{wd1,wd2}: 
\[
\nu_U = 2\nu_\phi - \nu_r, \quad 
\nu_L = 2(\nu_\phi - \nu_r).
\]

\end{itemize}

Fig.~\ref{fig5} presents the calculated correlations between the upper ($\nu_U$) and lower ($\nu_L$) QPO frequencies for Schwarzschild, RN, and ModMax black holes across the various QPO models, evaluated for different values of the ModMax parameter $\eta$. 
For the RN black hole, the electric charge is fixed at $Q = 0.5$. 
The plot also includes reference lines corresponding to characteristic frequency ratios of $3\!:\!2$, $4\!:\!3$, $5\!:\!4$, and $1\!:\!1$. 
The last ratio represents the special scenario where the twin QPO peaks coalesce into a single dominant frequency, often described as the \emph{``QPO graveyard''}.

\subsection{QPO Orbits}
\begin{figure*}[h!]
\centering
\includegraphics[width=0.32\linewidth]{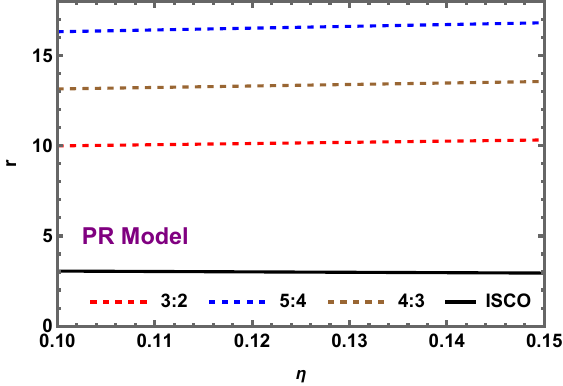}
\includegraphics[width=0.32\linewidth]{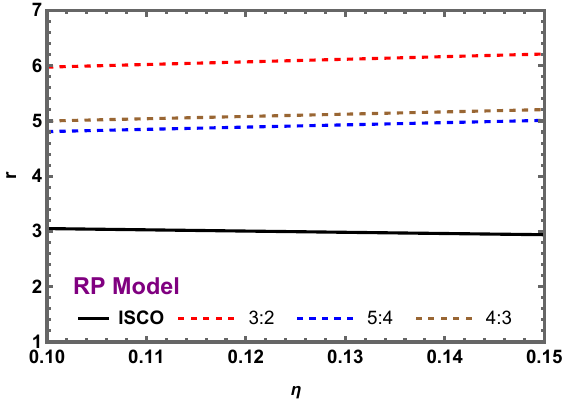}
\includegraphics[width=0.32\linewidth]{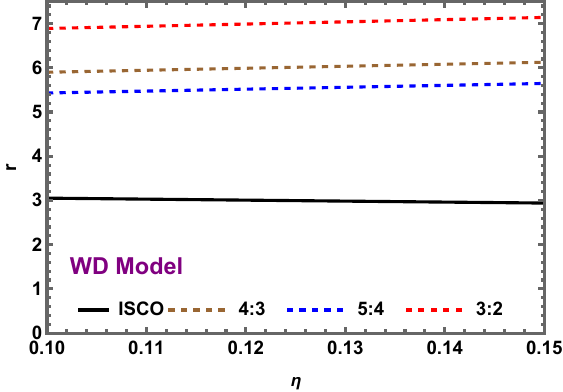}
\includegraphics[width=0.32\linewidth]{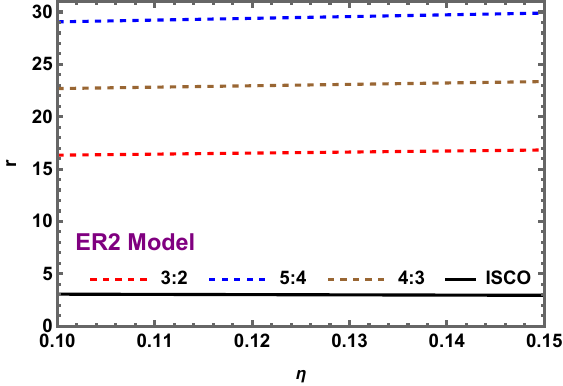}
\includegraphics[width=0.32\linewidth]{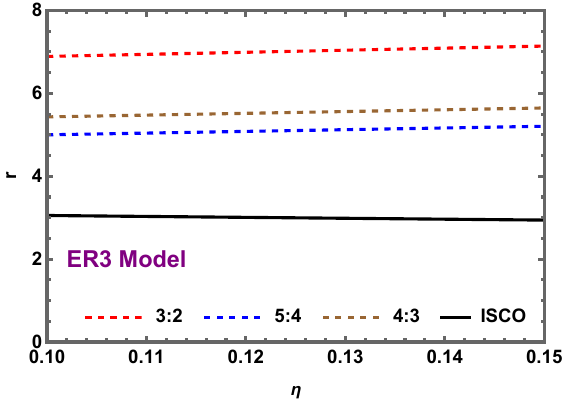}
\includegraphics[width=0.32\linewidth]{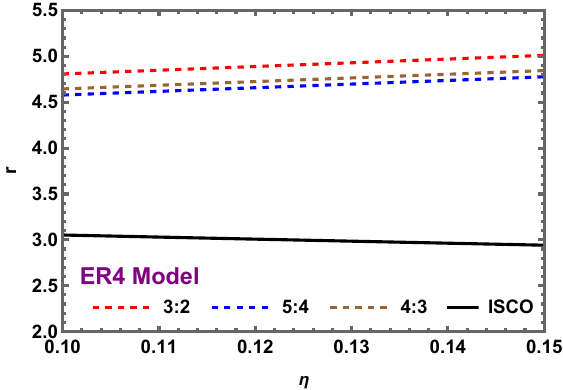}
\caption{Variation of  QPO orbit radii with the ModMax parameter $\eta$ for different QPO models} 
\label{fig6}
\end{figure*}

We now explore the dependence of the QPO orbital radii on the ModMax parameter $\eta$, focusing on locations where characteristic frequency ratios such as $3\!:\!2$, $4\!:\!3$, and $5\!:\!4$ emerge within the framework of the various QPO models considered. These resonant radii are obtained by solving the corresponding resonance condition,
\begin{equation}
\alpha\,\nu_U(M, r, Q, \eta) = \beta\,\nu_L(M, r, Q, \eta), \label{resonance}
\end{equation}
Here, $\alpha$ and $\beta$ denote integers specifying the resonance ratio. For the PR, RP, WD, and ER2–ER4 models, this resonance condition is solved numerically to determine the corresponding orbital radius $r$ for various values of the ModMax parameter $\eta$. The resulting numerical solutions are presented in Fig.~\ref{fig6}. We show the variation of the orbital radii with respect to the ModMax parameter $\eta$ for all the six QPO models considered in our study namely, PR, RP, WD, ER2, ER3, and ER4. Each panel illustrates the orbital radii of all the QPO models corresponding to the resonant frequency ratios $3\!:\!2$, $4\!:\!3$, and $5\!:\!4$, alongwith the innermost stable circular orbit (ISCO) indicated by a solid black line.

We observe that the ER2 model produces significantly larger QPO generating radii compared to the other models. The magnitude is followed by PR and WD models. On the other hand the ER4 model generates the smallest QPO orbital radius among the six models considered. Another point to note here is that the QPO orbital radii monotonically increases with increasing ModMax parameter $\eta$ corresponding to all the six models implying that as the deviation from standard electrodynamics becomes more pronounced, the corresponding resonance orbits shift outward, forming at larger radial distances from the black hole.

\section{Monte Carlo Markov chain (MCMC) analysis}
This section offers the Markov Chain Monte Carlo analysis done in order to constrain the ModMax black holes' parameters. We use the observational data of the following black hole sources across different mass regimes: GRO J1655–40, XTE J1550–564, GRS 1915+105, H 1743–322, M82 X-1, and Sgr A*. Within this sample, GRO J1655–40, XTE J1550–564, GRS 1915+105, and H 1743–322 are identified as stellar-mass black hole candidates, whereas M82 X-1 is classified as an intermediate-mass black hole source, and Sgr A* represents the supermassive black hole residing at the Galactic center. A concise summary of these black hole systems, together with their observed QPO properties, is presented in Table \ref{tab1}. In the present analysis, the relativistic precession (RP) model is employed as a representative framework to illustrate the implementation of Markov Chain Monte Carlo (MCMC) methods for constraining the parameters of black holes. The selection of this model is made purely for methodological demonstration, without implying any specific preference or physical superiority of the RP scenario.

We have the Bayesian posterior distribution given as:
\begin{equation}
P(\boldsymbol{\theta} | D, M) = \frac{P(D | \boldsymbol{\theta}, M)\, \pi(\boldsymbol{\theta} | M)}{P(D | M)},
\label{eq:posterior}
\end{equation}
with $\pi(\boldsymbol{\theta})$ being the prior distribution for the parameters $\boldsymbol{\theta} = \{M, \eta,\frac{Q}{M},  \frac{r}{M}\}$, and $P(D | \boldsymbol{\theta}, M)$ is the likelihood function. 

{\LARGE
\begin{table*}[htb]
\centering
\begin{tabular}{|c|c|c|c|}
\hline
\textbf{Source} & \textbf{Mass} (in $M_\odot$) & \textbf{Upper Frequency (Hz)} & \textbf{Lower Frequency (Hz)} \\
\hline
GRO J1655$-$40 & $5.4 \pm 0.3$ \cite{t57} & $441 \pm 2$ \cite{t58} & $298 \pm 4$ \cite{t58} \\
\hline
XTE J1550$-$564 & $9.1 \pm 0.61$ \cite{t59} & $276 \pm 3$ & $184 \pm 5$ \\
\hline
GRS 1915$+$105 & $12.4^{+2.0}_{-1.8}$ \cite{t60} & $168 \pm 3$ & $113 \pm 5$ \\
\hline
H 1743$+$322 & $8.0 - 14.07$ \cite{t61,t62,t63} & $242 \pm 3$ & $166 \pm 5$ \\
\hline
M82 X-1 & $415 \pm 63$ \cite{nature} & $5.07 \pm 0.06$\cite{nature}  & $3.32 \pm 0.06$\cite{nature}  \\
\hline
Sgr A$^*$ & $(3.5 - 4.9) \times 10^6$ \cite{t64,t65} & $(1.445 \pm 0.16) \times 10^{-3}$\cite{t66} & $(0.886 \pm 0.04) \times 10^{-3}$\cite{t66} \\
\hline
\end{tabular}
\caption{Observational QPO data for different Black hole sources with estimated mass  \cite{52}}
\label{tab1}
\end{table*}
}

Assuming Gaussian priors for each parameter, 
\begin{equation}
\pi(\theta_i) \propto \exp\left(-\frac{1}{2} \left(\frac{\theta_i - \theta_{0,i}}{\sigma_i} \right)^2\right), \quad \theta_{\mathrm{low},i} < \theta_i < \theta_{\mathrm{high},i},
\end{equation}
where $\theta_{0,i}$ and $\sigma_i$ represent the mean and standard deviation of the corresponding parameters. The log-likelihood function, which represents the natural logarithm of the likelihood function, can in the context of upper and lower QPO frequency be formulated as \begin{equation}
\log \mathcal{L} = \log \mathcal{L}_U + \log \mathcal{L}_L,
\end{equation}
where
\begin{equation}
\log \mathcal{L}_U = -\frac{1}{2} \sum_i \frac{(\nu^{\mathrm{obs}}_{\phi,i} - \nu^{\mathrm{th}}_{\phi,i})^2}{(\sigma^{\mathrm{obs}}_{\phi,i})^2},
\end{equation}
\begin{equation}
\log \mathcal{L}_L = -\frac{1}{2} \sum_i \frac{(\nu^{\mathrm{obs}}_{L,i} - \nu^{\mathrm{th}}_{L,i})^2}{(\sigma^{\mathrm{obs}}_{L,i})^2},
\end{equation}
Here, $\nu^{\mathrm{obs}}_{\phi,i}$ and $\nu^{\mathrm{obs}}_{L,i}$ denote the observed orbital and lower QPO frequencies, respectively, whereas $\nu^{\mathrm{th}}_{\phi,i}$ and $\nu^{\mathrm{th}}_{L,i}$ correspond to their theoretical counterparts computed within the framework of the RP model.

In this analysis, we employ the observed QPO data corresponding to the six black hole sources listed in Table~\ref{tab1}. Utilizing the available prior information, we generate $10^5$ samples for each parameter by adopting Gaussian priors, enabling a comprehensive exploration of the multidimensional parameter space. The objective is to determine the most likely values of the parameters $\{M,\, \eta,\, \frac{Q}{M},\, \frac{r}{M}\}$ that best reproduce the observational data.

\begin{figure*}[t]
    \centering
    \subfloat[GRO J1655-40]{%
        \includegraphics[width=0.45\textwidth]{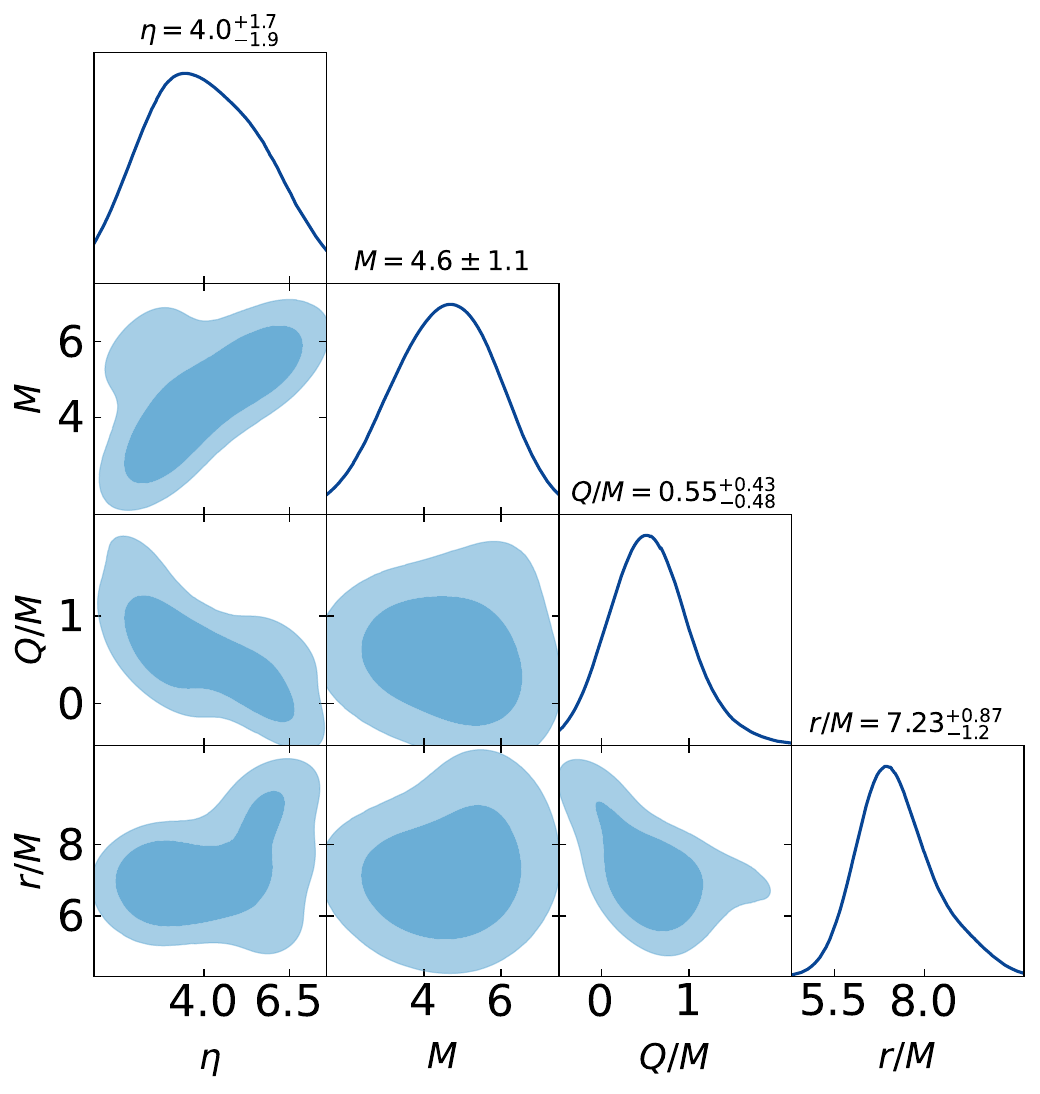}
    }\hfill
    \subfloat[XTE J1550-564]{%
        \includegraphics[width=0.45\textwidth]{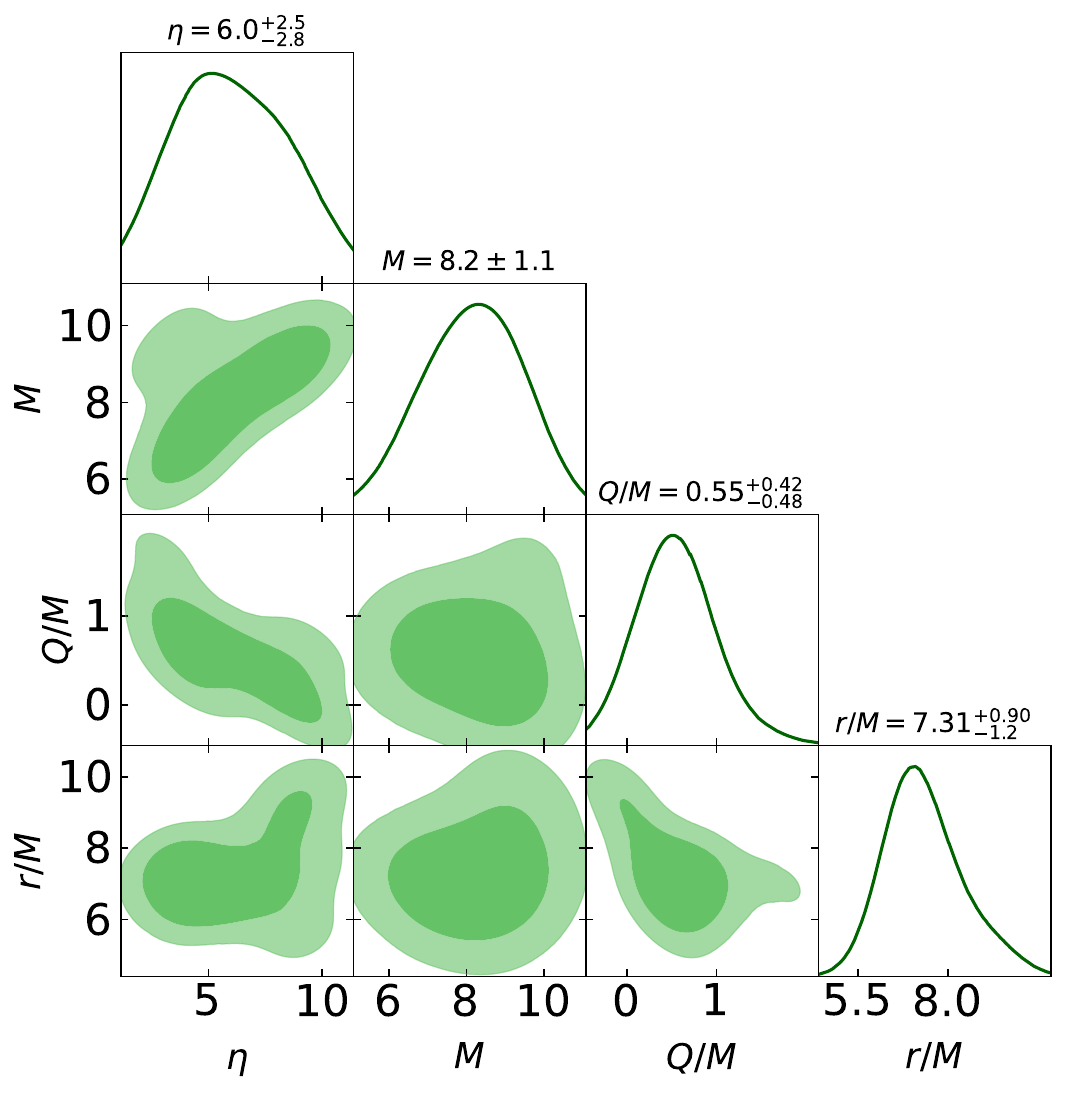}
    }\hfill
    \subfloat[GRS 1915+105]{%
        \includegraphics[width=0.45\textwidth]{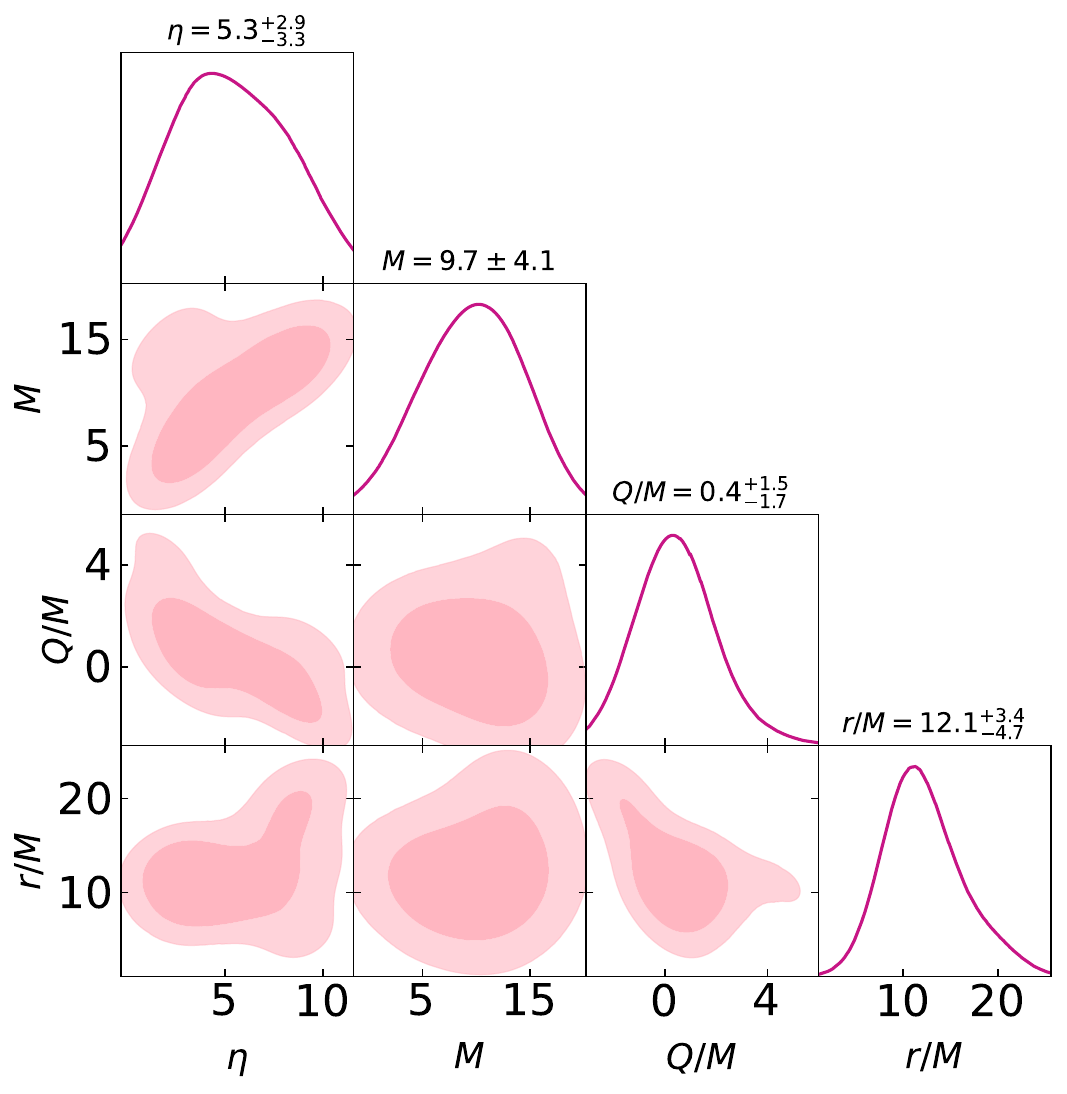}
    }\hfill
    \subfloat[H 1743+322]{%
        \includegraphics[width=0.45\textwidth]{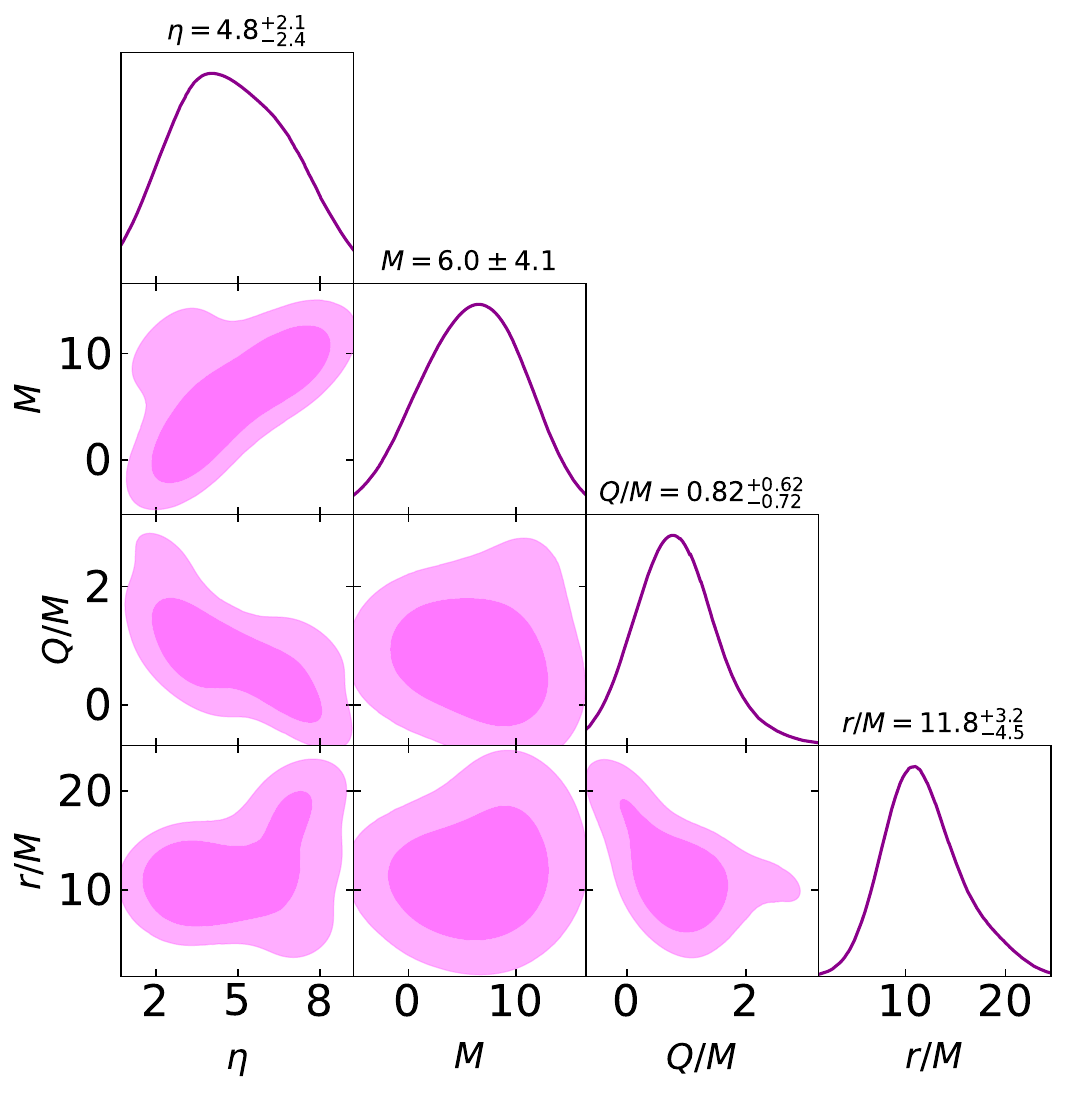}
    }\hfill
    \caption{Corner plots illustrating the posterior probability distributions of the parameters \(M\), \(\eta\), \(Q/M\), and \(r/M\) derived from the MCMC sampling for each of the analyzed black hole systems.
}
    \label{fig7}
\end{figure*}

\begin{figure*}[t]
    \centering
    \subfloat[M82 X-1]{%
        \includegraphics[width=0.45\textwidth]{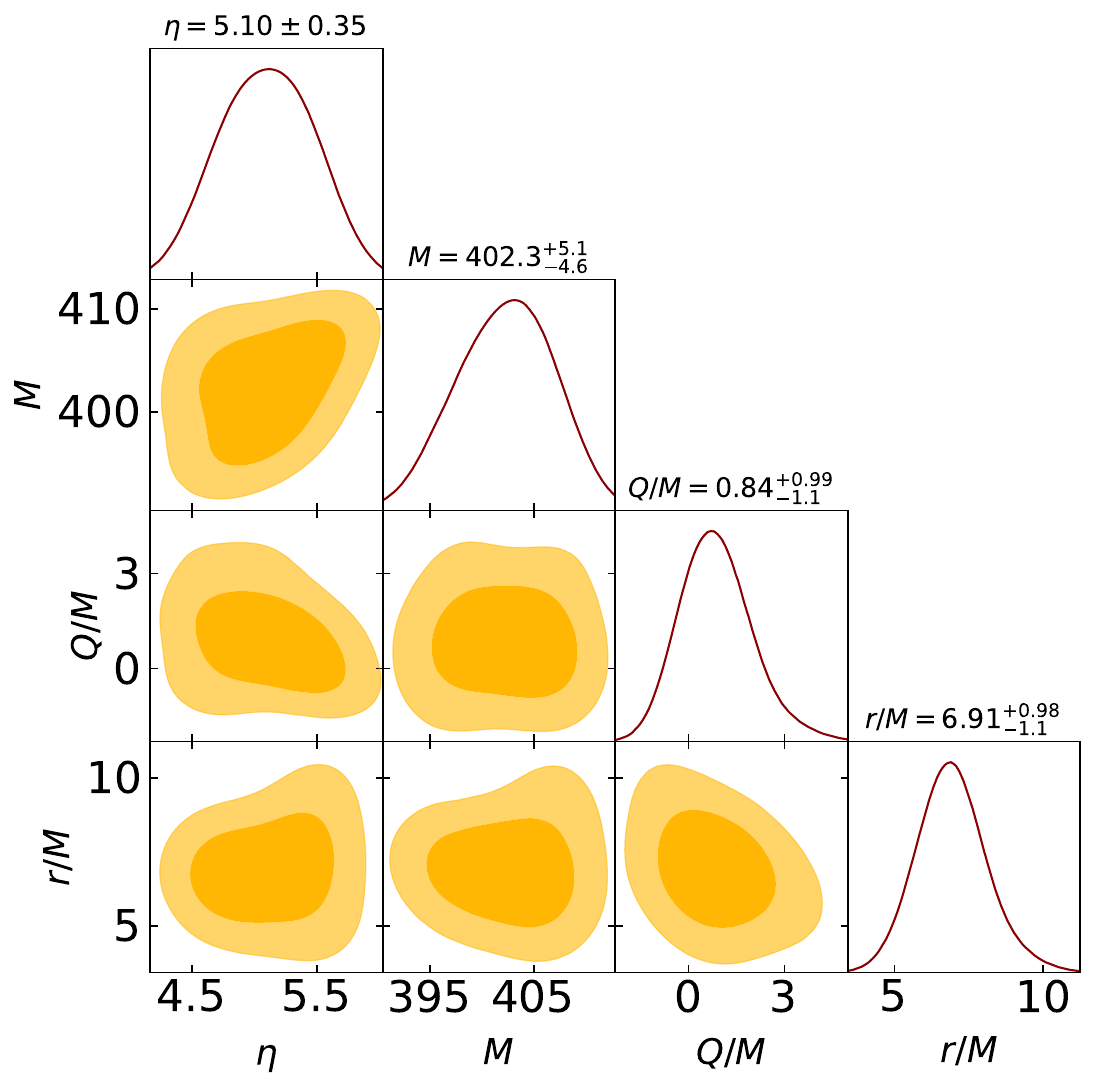}
    }\hfill
    \subfloat[Sgr A*]{%
        \includegraphics[width=0.45\textwidth]{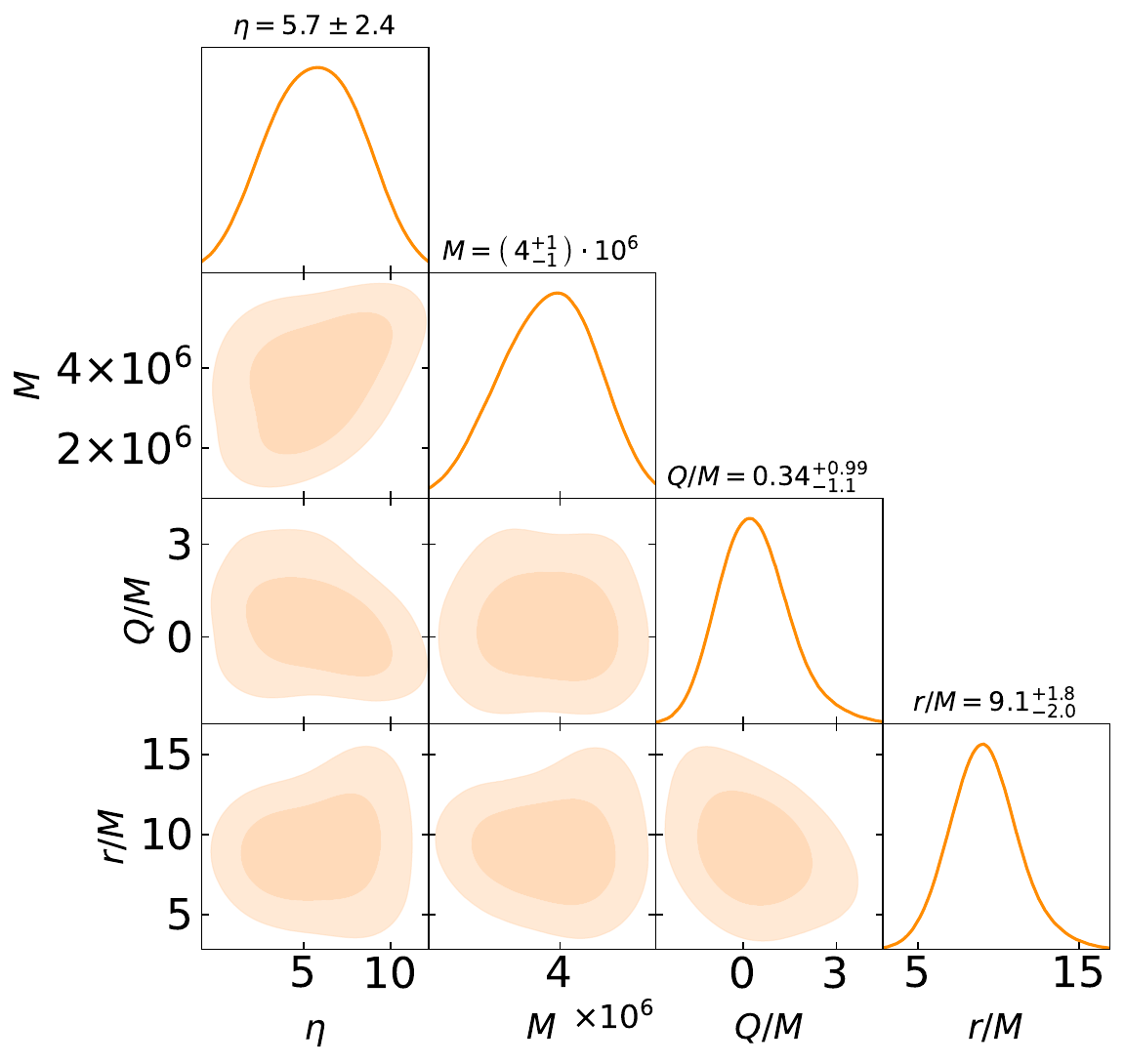}
    }
    \caption{Corner plots illustrating the posterior probability distributions of the parameters \(M\), \(\eta\), \(Q/M\), and \(r/M\) derived from the MCMC sampling for each of the analyzed black hole systems.}
    \label{fig8}
\end{figure*}

In Figure \ref{fig7} and \ref{fig8}, we present the corner plots obtained from the MCMC analysis. In this analysis, the corner plots display the joint posterior probability distributions of the model parameters, highlighting their corresponding confidence regions. The contours mark the \(68\%\) (\(1\sigma\)) and \(95\%\) (\(2\sigma\)) credible intervals, representing the most probable parameter ranges derived from the MCMC sampling. The darker shaded regions denote the \(1\sigma\) confidence level, while the lighter areas extend to the broader \(2\sigma\) uncertainty range. For the stellar-mass black holes, the MCMC analysis yields mass estimates consistent with observational data: \(M = 4.6 \pm 1.1\,M_\odot\) for GRO J1655–40, \(8.2 \pm 1.1\,M_\odot\) for XTE J1550–564, \(9.7 \pm 4.1\,M_\odot\) for GRS 1915+105, and \(6.0 \pm 4.1\,M_\odot\) for H 1743+322. The corresponding values of the ModMax parameter \(\eta\) are found to lie in the range \(4.0 \lesssim \eta \lesssim 6.0\), while the charge-to-mass ratio \(Q/M\) varies between 0.55 and 0.82. The dimensionless orbital radius parameter \(r/M\) is distributed around 7--12 for these systems, indicating relatively extended orbital configurations compared to the standard Maxwellian case. For the intermediate-mass black hole M82 X-1, we have the constraints on mass as \(M = 402^{+5.1}_{-4.6}\,M_\odot\), on $\eta$ as  \(\eta = 5.10 \pm 0.35\), on the charge-to-mass ratio of \(Q/M = 0.84^{+0.99}_{-1.1}\), and on the orbital radius as \(r/M = 6.91^{+0.98}_{-1.1}\)

The posterior distributions suggest that the ModMax parameter \(\eta\) remains of order unity across all black hole sources, implying a modest but consistent deviation from conventional electrodynamics. The relatively constrained values of \(\eta\) and \(Q/M\) across distinct mass scales indicate that nonlinear electromagnetic effects introduced by the ModMax framework could influence the quasi-periodic oscillation frequencies. The relatively large $Q/M$ ratios inferred in this analysis are difficult to reconcile with standard astrophysical expectations, since any macroscopic charge on a black hole would be rapidly neutralized by the surrounding plasma. We therefore suggest that the charge parameter in the ModMax framework should be interpreted as an effective parameter, rather than as a literal astrophysical electromagnetic charge. From this perspective, the inferred charge may instead encode additional degrees of freedom arising from nonlinear electrodynamic effects.

We summarize the results in from the MCMC analysis in Table \ref{tab2}. We observe that the constraints on the mass $M$ is in good agreement with the observational values as tabulated in \ref{tab1}. The constrained values of \(\eta\), \(Q/M\) and \(r/M\) are also shown in Table \ref{tab2}

\begin{table*}[htb]
\centering
\setlength{\tabcolsep}{12pt} % wider column spacing
\renewcommand{\arraystretch}{1.4} % taller rows
\begin{tabular}{|c|c|c|c|c|}
\hline
\textbf{Source} & \boldmath{$M$} & \boldmath{$\eta$} & \boldmath{$Q/M$} & \boldmath{$r/M$} \\
\hline
  \textbf{GRO J1655-40}     & $4.6 \pm 1.1$                  & $4.0^{+1.7}_{-1.9}$ & $0.55^{+0.43}_{-0.48}$ & $7.23^{+0.87}_{-1.2}$ \\
\hline
\textbf{XTE J1550-564}   & $8.2 \pm 1.1$                    & $6.0^{+2.5}_{-2.8}$ & $0.55^{+0.42}_{-0.48}$        & $7.31^{+0.90}_{-1.2}$ \\

\hline
 \textbf{GRS 1915+105}   & $9.7\pm 4.1$          & $5.3^{+2.9}_{-3.3}$        & $0.4^{+1.5}_{-1.7}$ & $12.1^{+3.4}_{-4.7}$ \\
\hline
\textbf{H 1743+322}     & $6.0\pm 4.1$                   & $4.8^{+2.1}_{-2.4}$ & $0.82^{+0.62}_{-0.72}$ & $11.8^{+3.2}_{-4.5}$ \\
\hline
\textbf{M82 X-1}  & $402.3^{+5.1}_{-4.6}$  & $5.10 \pm 0.35$    & $0.84^{+0.99}_{-1.1}$ & $6.91^{+0.98}_{-1.1}$ \\
\hline
\textbf{Sgr A*}  & $(4^{+1}_{-1}) \times 10^6$ & $5.7 \pm 2.4$    & $0.34^{+0.99}_{-1.1}$ & $9.1^{+1.8}_{-2.0}$ \\
\hline

\end{tabular}
\caption{Posterior constraints on the parameters \(M\), \(\eta\), \(Q/M\), and \(r/M\) derived from the MCMC parameter estimation.}
\label{tab2}
\end{table*}

\section{conclusion and discussion}
In this work, we have investigated the motion of neutral test particles and the associated quasi-periodic oscillations (QPOs) in the spacetime of a ModMax black hole, a nonlinear extension of the standard Maxwell theory coupled to general relativity. The analysis focused on examining how the ModMax parameter \(\eta\) influences the orbital structure, fundamental frequencies, and observable QPO features. Our results demonstrate that the ModMax parameter \(\eta\) plays a crucial role in modifying the effective potential, orbital frequencies, and resonant radii, thereby leaving measurable imprints on the QPO spectrum compared to the Reissner-Nordström and Schwarzschild cases. We analyzed the effective potential for neutral test particle and observed that as $\eta \to 0$ the effective potential resembles that of the RN black hole case and as $\eta \to \infty$ it resembles that of the Schwarzschild case.

We then considered particles in circular motion and analyzed their specific energy and angular momentum and observed that both quantities decrease as the ModMax parameter \(\eta\) increases. This trend indicates that the orbits become more tightly bound, leading to an outward shift of the stable circular trajectories. Nevertheless, the pronounced reduction in energy and angular momentum can also diminish the range of orbital stability, rendering certain trajectories more prone to instability. In the extremal limits of \(\eta\), we again observe similar behaviour as seen in the effective potential.

Subsequently, we investigated the Innermost Stable Circular Orbit (ISCO). Our analysis revealed that, for any given charge \(Q\), the ISCO radius expands progressively as the ModMax parameter \(\eta\) increases, suggesting that nonlinear electromagnetic corrections drive the stable orbits farther from the black hole. Moreover, for fixed \(\eta\), the ISCO radius exhibits an almost linear growth with respect to the charge parameter \(Q\), reflecting the combined influence of charge and nonlinear electrodynamic effects on orbital stability. Furthermore, the analysis of the Keplerian motion of test particles revealed that the orbital frequency \(\Omega_{\phi}\) diminishes with increasing radial distance. The inclusion of nonlinear electrodynamic effects further suppresses \(\Omega_{\phi}\) relative to the Schwarzschild limit. As the ModMax parameter \(\eta\) grows, this reduction becomes more pronounced, underscoring its influence on orbital dynamics. In the large and small \(\eta\) regime, the frequency profiles smoothly converge toward those characteristic of the Schwarzschild and RN geometry respectively, reflecting a continuous transition from nonlinear to linear electrodynamics.

We extended our analysis of the QPOs by investigating the epicyclic motions. We consider all six QPO models namely PR, RP, WD, ER2, ER3, and ER4. Our comparative analysis of different QPO models revealed distinct trends in the locations of the resonance orbits. Among the examined frameworks, the ER2 model yielded the largest QPO radii, followed by the PR and WD models, while the ER4 model predicted the smallest orbital radii. Across all cases, the QPO-generating radii were found to increase monotonically with the ModMax parameter \(\eta\), indicating that stronger deviations from linear electrodynamics tend to shift the resonance orbits outward. This behaviour highlights the sensitivity of QPO formation zones to nonlinear electromagnetic corrections in the underlying spacetime geometry.

Finally, to connect our theoretical framework with observations, we employed a Markov Chain Monte Carlo (MCMC) analysis to constrain the parameters \(\{M, \eta, Q/M, r/M\}\) of the ModMax black hole model using QPO data from six well-established black hole sources. These include four stellar-mass systems (GRO J1655–40, XTE J1550–564, GRS 1915+105, and H 1743–322), the intermediate-mass black hole M82 X-1, and the supermassive black hole Sgr A*, thereby covering a broad range of mass scales. In this work, we restrict our analysis to the RP model solely for the purpose of demonstrating the methodology. Consequently, the parameter constraints obtained and the physical interpretations inferred from the MCMC analysis are specific to the RP model considered here. Different QPO models may lead to quantitatively different constraints and potentially distinct physical interpretations, and exploring such model dependence is left for future investigations.\\
The results of the MCMC analysis are tabulated in Table \ref{tab2}, providing the constrained values of the above mentioned parameters. It is observed that the constrained mass value is in good agreement with the available observational data which is given in Table \ref{tab1}. The MCMC analysis also suggest that the ModMax parameter \(\eta\) is of the order unity implying a modest but consistent deviation from conventional electrodynamics. \\

It is worth noting that, although the observational data used in our study most likely correspond to rotating black holes, our analysis intentionally omits the spin parameter to isolate the intrinsic effects of nonlinear electrodynamics on QPO frequencies. However, the deliberate omission of rotation inevitably introduces a degeneracy in the parameter estimation. In particular, the inferred constraints on the nonlinear electrodynamic parameter $\eta$ may effectively absorb frequency shifts that would otherwise be attributed to black hole spin. As a result, although the primary objective of this manuscript is to examine whether nonlinear electrodynamic effects alone can reproduce frequency variations commonly associated with rotation, it should be emphasized that the best-fit values of the NED parameter obtained from the MCMC analysis are unlikely to arise purely from nonlinear effects. Instead, they should be interpreted as effective parameters that may implicitly encode, at least in part, the influence of rotational dynamics. Our findings suggest that a moderate nonlinear charge can phenomenologically accommodate the observed QPO frequencies within the present model framework.\\

Next, we compare our results obtained within the RP model with those of Ref.~\cite{bidyut}, in which a different nonlinear electrodynamic model was investigated in the context of observed QPOs. While the two studies differ in the underlying NLED formulation, a meaningful comparison can be made at the level of inferred parameter trends and effective physical interpretations.  In our ModMax-based analysis, the nonlinear electrodynamic effects are characterized by the parameter $\eta$, whereas in the comparative study the corresponding role is played by the parameter $b$.  Despite this difference in parametrization, both analyses yield broadly consistent mass estimates for the considered sources, which are in good agreement with independent astrophysical constraints.  A comparison of the $r/M$, reveals that both models predict the value of $r/M$ within $r/M \sim 5$--$12$. While the exact numerical values differ between the two studies, the overall scale and ordering across sources remain comparable.  In both models, the inferred values of the charge-to-mass ratio $Q/M$ are found to be close to unity.  Although the corresponding $Q/M$ values obtained in Ref.~\cite{bidyut} are somewhat lower, they remain of the same order of magnitude.  Such large values are difficult to interpret as a literal astrophysical electric charge, since any real charge would be rapidly neutralized in an astrophysical environment. This supports the interpretation of  $Q/M$   as an effective parameter that captures the influence of nonlinear electrodynamic effects, rather than representing a true electromagnetic charge of the black hole.  Similarly, although both $\eta$ and $b$ act as measures of deviation from the Maxwell limit, their posterior distributions are not directly comparable on a one-to-one basis. Instead, they should be viewed as model-specific parameters whose numerical values depend on the particular functional form of the nonlinear electrodynamic Lagrangian.  Despite these differences, both parameters consistently take nonzero values greater than unity across multiple sources, suggesting that nonlinear electromagnetic effects may play a phenomenologically relevant role in shaping QPO frequencies within the RP model.  Overall, this comparison highlights that certain inferred quantities, such as the black hole mass and the characteristic QPO radius, remain relatively robust across different nonlinear electrodynamic models. The constraints on the effective charge-to-mass ratio are found to be of order unity in both models, although slightly lower values are obtained in Ref.~\cite{bidyut}. In contrast, while the nonlinear electrodynamic parameters in the two models are of the same order of magnitude, $\mathcal{O}(10^{0})$, their numerical values differ significantly. This model dependence shows the importance of exercising caution when attributing direct physical meaning to individual parameter values and motivates future studies exploring multiple QPO prescriptions and nonlinear electrodynamic theories within a unified statistical framework.

\end{document}